\documentclass[12pt]{article}
\usepackage{graphicx}
\usepackage{amssymb,amsmath,amsfonts,palatino,amsthm}
\usepackage{amssymb}
\usepackage{epstopdf}
\newcommand{\f}{\begin{equation}}
\newcommand{\ff}{\end{equation}}
\newcommand{\blankline}{\vskip .3cm}
\begin{document}

\section*{A naturalist account of the limited, and hence reasonable, effectiveness of mathematics in physics}

\centerline{\it Lee Smolin}

\blankline

My aim in this essay is to propose a conception of mathematics that is fully consonant with {\it naturalism.}  By that I mean the hypothesis that {\it everything that exists is part of the natural world, which makes up a unitary whole.}  This is in contradiction with the Platonic view of mathematics held by many physicists and mathematicians according to which, {\it mathematical truths are facts about mathematical objects which exist in a separate, timeless realm of reality, which exists apart from and in addition to physical reality.}  

If you are a Platonist the question of the relationship between mathematical truths and true facts about nature concerns a hypothetical correspondence between two realms of existence, physical reality and the Platonic realm of mathematical reality.  The effectiveness of mathematics in physics is in this context   mysterious because proponents of this view have failed to explain both how there could be such a correspondence and how we, as beings trapped in time bound physical reality, can have certain knowledge of the hypothesized separate realm of mathematical reality\footnote{One radically simple approach to bridging this corespondence is the mathematical universe hypothesis\cite{MUH}.}.  

If you are a Platonist, then I would argue you cannot be a naturalist, because you believe that something exists apart from physical reality.  At best you are a dualist.  Similarly, if you believe that the ultimate goal  of physics is to discover a mathematical object, $\cal O$, which is in perfect correspondence with nature, such that every  true fact about the universe, or its history, is isomorphic to a true fact about $\cal O$, then you are also not a naturalist because you not only believe in the existence of something which is not part of nature, you believe that everything that is true about nature is explained by a true fact about something which exists apart from nature.  You are instead a kind of mystic, believing in the prophetic power of the study of something which exists outside of time and apart from nature.


If, on the other hand, all that exists is physical reality then mathematical knowledge must be an aspect of knowledge about physical reality.  The view I will propose answers Wigner's query about the "{\it unreasonable} effectiveness of mathematics in physics" by showing that the role of mathematics within physics is {\it reasonable}, because it is limited.  In particular, there is no mathematical object which is isomorphic to the universe as a whole, and hence no perfect correspondence between nature and mathematics.  There indeed may be properties enjoyed by physical reality which have no counterpart in mathematics.
I will mention two below.  

Mathematics thus has no prophetic role in physics, which would allow us an end run around the hard slog of hypothesizing physical principles and theories and testing their consequences against experiment.  Mathematical beauty may inspire us, while remaining, like most sources of beauty, apt to mislead us if intoxication becomes a substitute for thought.  Mathematics is nonetheless the second most important tool in our physicists' toolkit, after the methods of experiment and observation.  Even if not every fact about the natural world has a correspondence with a property of a mathematical object, it is nonetheless important  that many experimentally accessible facts about the natural world can be predicted and explained by a process that involves modelling a physical experiment in mathematical terms.  So even a limited and partial correspondence between mathematical results and the results of experiments  challenges our attempts to account for mathematics completely within a naturalist perspective.  

At the same time attempts to formulate a non-Platonic foundations of mathematics have been unsatisfactory, as they too fail to account for the effectiveness of mathematics as a tool of physics.  If mathematical truths are merely truths by convention or agreement, how could it possibly be the case that they sometimes correspond to facts about nature?

So a new conception of mathematics is needed which is entirely naturalist and regards mathematical truths as truths about nature.  In this essay I sketch a proposal for such a view. The key it turns out is the conception of time.  I propose that to get a conception of mathematics within naturalism it is essential to regard time as an essential aspect of nature, in a sense to be specified shortly.  I thus propose to call this new
\footnote{Related explorations of the notion of naturalism are discussed in 
\cite{me-brick,Jaron-gadget,GS}.} view, {\it temporal naturalism\cite{TN}.} 

This  view which has developed from an eight year collaboration with Roberto Mangabeira Unger.  It is presented more fully in our recently published book\footnote{
Mangabeira Unger also will have an essay in this competition\cite{RMU-FQXi} which presents a view complementary to this one.}, from which the following is taken\cite{SURT}.  

Mangabeira Unger and I hypothesize two principles which we take to define temporal naturalism.

\begin{enumerate}

\item{} {\bf The singular universe:}  All that exists is part of a single, causally connected universe.  The universe and its history have no copies, and are not part of any ensemble.  There is no other mode of existence, in particular neither a Platonic realm of mathematical objects nor an ensemble of possible worlds exist apart from the single universe.  

\item{}{\bf The inclusive reality of time:}  All that is real or true is such within a moment, which is one of a succession of moments.  The activity of time is a process by which novel events are generated out of a presently existing, thick set of present events.  There are no eternal laws; laws are subsidiary to time and to a fundamental activity of causation and may evolve.  There is an objective distinction between past, present and future.

\end{enumerate}

These principle are really nothing but a deepening of what it means to be a naturalist.  If all that exists is part of nature, then all chains of causation must refer back only to past natural events.  Moreover we adopt a strong form of Einstein's principle of no unreciprocated action according to which there can be no entity A which plays a role in explaining an event B, that cannot itself be influenced by prior physical events.  Among the things that violate a strict definition of naturalism are explanations that refer to ideal, timeless absolute elements such as absolute space and time, non-dynamical fixed background geometries, whether of spacetime or Hilbert space, or absolute, timeless laws, which are imagined to act to cause things to happen in nature but are never themselves acted upon.  

In the following I sketch a view of mathematics consistent with these principles within which the necessary, but necessarily limited role of mathematics in physics is reasonable.  

\section*{A new conception: mathematics as a study of evoked reality}



According to the view most commonly held among physicists and mathematicians, mathematics is the study of a timeless but real realm of mathematical objects.  This contradicts our principles twice over, both because there is no real realm other than our one universe and second because there is nothing real or true that is timeless.  


The choice between whether mathematics is discovered or invented is a false choice. Discovered implies something already exists and it also implies we have no choice about what we find. Invented means did not exist before AND we have choice about what we invent. 

So these are not opposites. These are two out of four possibilities on a square whose dimensions are whether an entity allows choice or not and whether it already exists or not. 

I would like to propose that there is a class of facts about the world, which concerns structures and objects which come to exist at specific moments, which, nevertheless,  have rigid properties once they exist.  

Let us call this possibility {\it evoked.}  I would then propose that mathematics consists of the study of certain of these evoked structures. The four possibilities are indicated in the following diagram:
\[
\begin{array}{cccc}
&  Existed \  prior?          &		Yes        		& 	No \\
Has  \ rigid     \  properties?  & & & \\
Yes             &	&	Discovered   	&	Evoked   \\
No.              &	&	Fantasized     	&	Invented   \\
\end{array}
\]

There are many things that did not exist before we bring them into existence but about which we have no choice, or only  highly constrained choices, once it does exist. So the notion of evocation applies to many things besides mathematics. 

For example, there are an infinite number of games we might invent. We invent the rules but, once invented, there is a set of possible plays of the game which the rules allow.   We can explore the space of possible games by playing them, and we can also in some cases deduce general theorems about the outcomes of games. 

It feels like we are exploring a pre-existing territory as we often have little or no choice, because there are often surprises and incredibly beautiful insights into the structure of the game we created. But there is no reason to think that game existed before we invented the rules. What could that even mean?

There are many other classes of things that are evoked. There are forms of poetry and music that have rigid rules which define vast or countably infinite sets of possible realizations. They were invented, it is absurd to think that haiku or the blues existed before particular people made the first one. Once defined there are many discoveries to be made exploring the landscape of possible realizations of the rules. A master may experience the senses of discovery, beauty and wonder, but these are not arguments for the prior or timeless existence of the art form independent of human creativity. 

It just happens to be a true fact about the world that it is possible to invent novel games, or forms which, once brought into existence, have constraints or rules which define a vast or infinite space of realizations. 

When a game like chess is invented a whole bundle of facts become demonstrable, some of which are theorems that become provable through straightforward mathematical reasoning.  As we do not believe in timeless Platonic realities, we do not want to say that chess always existed-in our view of the world, chess came into existence at the moment the rules were codified.  This means we have to say that all the facts about it became not only demonstrable, but true, at that moment as well.  Our time bound world is just like that: there are things that spring into existence, along with a large and sometimes even infinite set of true properties.  This is what the word evoked means to convey: the facts about chess are evoked into existence by the invention of the game.

The concept of evoked truths depends essentially on the reality of time because it has built into it the distinction between past, present and future.

Once evoked, the facts about chess are objective, in that if any one person can demonstrate them, any one can.  And they are independent of time or particular context, so long as it is after the invention of chess makes it possible to discuss them.  Furthermore, these facts are facts about our one world, just the same as facts about how many legs some insect has or which species can fly.  The latter facts were evoked by evolution acting through natural selection, the facts about chess were evoked by the invention of the game as a step in the evolution of human culture.

One consequence of the Platonic view is to deny the possibility of novelty.  No game, construction  or theorem is ever new because anything that humans discover or invent  existed already timelessly in the Platonic realm.  The alternative to believing in the timeless reality of any logically possible  game or species is believing in the reality of novelty.  Things come into existence and facts become true all the time.  This is one meaning of the reality of time.  Nature has within it the capacity to create kinds of events, or processes or forms which have no prior precedent.  We human beings can partake of this ability by the evokation of novel games and mathematical systems.

So it is not just human beings that have the power to evoke novel structures, which bring along with them novel facts.  Nature has this capacity as well and uses it on a range of scales from the emergence of novel phenomena which are describe by novel laws to the emergence of novel biological species which play novel games to dominate novel niches.

The notion of novel patterns or games evoked into existence gives a precise meaning to the concept of emergence.  In a timeless world 
emergence is always at best an approximate and inessential description because one can always descend to the timeless fundamental level of description according to which all that happens is the rearrangement of particles with timeless properties under timeless laws.  But once we admit the actuality of the emergence of novel games and structures with evoked properties, emergence has a fundamental, irreducible meaning.  

In fact, biological evolution proceeds by a sequence of evokings of novel 
games and structures.
Once cells with DNA and the standard biochemistry come into existence a vast landscape of possible species and ecologies opens up. As the biosphere evolves it discovers many niches where species may thrive. New innovations appear from time to time like eukaryotic cells, multi-cellularity, oxygen breathing, plants etc, which define further constraints which in turn make possible new variations, niches and innovations. All this is truly a wonder but it would add nothing and explain nothing to posit that there is a timeless platonic world of possible DNA sequences, species, niches, ecologies that are being realized. Such a belief would explain nothing about how the real biosphere evolves and raise many additional questions whose answers, if they had answers, would add nothing to our understanding of the history of life or allow us to predict features of future life any more that we already can. 

What applies to biology also applies to mathematics.  There is a potential infinity of FAS's (formal axiomatic systems). Once one is evoked it can be explored and there are many discoveries to be made about it. But that statement does not imply that it, or all the infinite number of possible formal axiomatic systems. existed before they were evoked. 

Indeed, its hard to think what belief in the prior existence of a FAS would add. Once evoked, some FAS have an infinite number of properties which can be discovered or proved, about which there is no choice. 
But the postulate that this FAS existed before being evoked would not  explain the existence of those true properties because it involves belief in something that itself needs explanation. If the FAS existed prior or timelessly, what brought it into existence? How can something exist now but also exist timelessly.  For if it only existed "outside time", would could we, who are time-bound, and only come into contact with other things that live in time, ever know of it? How can something exist and not be made of matter? How can something that is not made of matter be known about, explored or influence us, who are made of matter?

So the postulation of prior or timeless existence explains nothing that is not explained by the notion of being evoked. It raises several questions including the ones just mentioned that are even more difficult to answer, and which centuries of attempts by very bright people have not answered. 

Since the notion of evocation is sufficient to explain why a FAS once evoked has rigid properties to be discovered, the notion of prior or timeless existence is not helpful. And it requires us to believe in a whole class of existence, as well as belief in the existence of an infinite number of FAS's, for which there is no evidence. By Occam's razor, this is not plausible. 

Barry Mazur, in a very helpful essay\cite{BMazur}, asserts that any answer to Platonism has to say something about the nature of proof. First of all, proof is a specialization of rational argument. It is a true fact about the world of possibilities brought into being when we humans evolved that in many situations we can rely on rational argument to lead to unambiguous conclusions. It happens to be the case that there are classes of questions that can be decided unambiguously by rational argument based on public evidence. 

This fact of the reliability of rational deduction cannot be explained by pointing to an imagined world of timeless but existing logical forms, as that would raise more unanswerable questions of the above kind. So it has to be taken as a simple brute fact about the world that experience has long validated. 

The process of rational deduction has itself been formalized, so rational argument from evidence is also a formal game whose rules have been defined in a way that in some classes of questions defines constraints sufficient to lead to unambiguous conclusions. 

Among these classes are mathematical systems defined by rules, or FAS's. Proofs are first of all just instances of rational argument applied to FAS's to deduce true properties of them. Once evoked a FAS has many, often an infinite number of true properties that can be so established. 

Proofs can be formalized, and there may be different ways of doing that. Each formalization is itself a formal game which is evoked, after which it can also be studied and explored. One can then raise and answer questions about how different formal methods of proof are related to each other. 

The bottom line is this: we have a choice between simple wonder and mystification. We can wonder at the vast complexity and beauty that is created by novel games, ideas, formal systems, etc when they are evoked. That there is such possibility of novel systems to explore is a true fact about the world we find ourselves in, which is properly a source of wonder. 

Or we can make mystical pronouncements that attempt to explain the infinite possibilities that might be evoked by imagining they all exist in a timeless reality apart from what we see physically exists around us. But these mystical beliefs add nothing and explain nothing and indeed, as indicated above, involve us in a pile of questions that, unlike questions about mathematics, cannot be answered by rational argument from public evidence. Moreover, to assert that one's avocation is the exploration of some timeless unphysical reality is presumptuous and seems like a claim to special knowledge or authority that, in fact, contradicts the fact that mathematical arguments are just finely disciplined cases of the usual rational thinking that all humans constantly engage in to understand their world. 

Honest wonder about our world seems a better stance than mysticism, especially when what is involved is the highest form of rational creativity. For that reason it seems better to believe in the possibility of evocation to create novel realms of truth to be explored that did not exist before than to believe in a special ability to gain knowledge about a timeless realm disconnected from physical existence.

\subsection*{The reasonable effectiveness of mathematics in physics}

So the answer to Wigner's question is that mathematics is reasonably effective in physics, which is to say that, where ever it is effective,  there is reason for it.  But mathematics does not of itself lead to discoveries about nature, nor is physics the search for a mathematical object isomorphic to the world or its history.  There will never be discovered a mathematical object whose study can render unnecessary the experimental study of nature; there is no mathematical discovery in our future that will render moot from then on the experimental and observational basis of science.  It will always be the case that the use of mathematics to model nature will be partial-because no mathematical object is a perfect match for nature.  The use of mathematics in nature also  involves a large degree of arbitrariness, because those mathematical objects that provide partial mirrors of parts of the world are a small, finite subset of the potentially infinite number of mathematical objects that might be evoked.   So the effectiveness of mathematics in physics is limited to what is reasonable.  

Moreover, any view about the role of mathematics in physics has to deal with the troubling issue of underdetermination of the choice of mathematical models of physical systems.   Most mathematical laws used in physics do not uniquely model the phenomena they describe.  In most cases the equation describing the law could be complicated by the addition of extra terms, consistent with the symmetries and principles expressed, whose effects are merely too small too measure given state of the art technology.   These ``correction terms" may be ignored because they don't measurably affect the predictions, but only complicate the analysis.  That this is the right thing to do methodologicaly does not, however, change the fact, that every one of the famous equations we use is merely the simplest of a bundle of possible forms of the laws which express the same ideas, symmetries and principles, and have the same empirical content.

This fact of under-determination is a real problem for those views which assert that nature is mathematical or that there is a mathematical object which is an exact mirror of nature, for only one out of the bundle of equations can be the true reality or mirror.   Often we assert that the right one is the simplest, evoking a necessarily mystical faith in ``the simplicity of nature."  The problem is that it never turns out to be the case that the simplest version of a law is the right one.  If we wait long enough we always discover that the simplest version is in fact wrong, because the theory is superseded by a new theory.   The old equation turns out still to hold approximately, but with corrections which take a form that could not have been guessed or anticipated prior to the invention of the new theory.

Thus, Newton's laws were found to be corrected by terms from special relativity, and then corrected again by terms from general relativity.  Maxwell's equations received corrections that describe light scattering from light-a quantum effect that could have been modelled-but never anticipated-by Maxwell.  And so on.

The radical under determinacy of the mathematical representation of physics is however no problem for the view proposed here.  It is rather exactly what you would expect, if mathematics is a powerful tool for modelling data and discovering approximate and ultimately temporary regularities which emerge from large amalgamations of elementary unique events.
In this context we use the simplest equation that expresses a law, not because we believe nature is simple but because it is a convenience for us-it makes a better tool, much as a hammer with a handle moulded  to the hand is a better tool,
Moreover in this context every theory is an effective theory which means that the limitations on the domain of applicability are always explicit and the correction terms are always there and ready to be exploited when a boundary of the domain of applicability is approached.

\subsection*{The reasonable effectiveness of mathematics in mathematics}

A satisfactory view of mathematics must also explain the unreasonable effectiveness of mathematics in mathematics itself.  Why do developments in the elaboration of one core concept-say number, so often turn out to yield insight into another-say geometry?  Why does algebra turn out to be so powerful a tool in the study of topology?   Why do the different division algebras organize the classification of the possible symmetry groups of continuous geometries?   If mathematics is just the free exploration of arbitrary ideas and axiom systems, why do these explorations so often intersect, and why are these intersections so productive of insight?

A short answer is that the contents of mathematics is far from arbitrary-while an infinite number of mathematical objects might potentially be envoked- the few that prove interesting develop a very small number of core concepts. These core concepts  are not arbitrary-they are elaborations of structures which are discovered during the study of nature. 

There are four of these core concepts:  number, geometry, algebra and logic.  They each capture a key aspect of the world and our interaction with it.  Number captures the fact that the world contains distinguishable objects which can be counted.  Geometry captures the fact that objects are found to take up space and form shapes.  Algebra captures the fact that objects and number can be transformed, by processes carried out in time.  And logic is the distillation of the fact that we can reason about the first three concepts, and so deduce predictions for future observations from properties of past observations. 

The bulk of mathematics consists of elaborations of these four core concepts.  In the course of these elaborations we often find developments of one bear on another.  These intersections tell us that these concepts go back to nature, which is a unity.  For example, the elaboration of the concepts of space and number  often intersect because space and number are both features of nature and hence are highly inter-related from the start.  Hence, the discovery that a relation among numbers represents or is isomorphic to a relation amongst another strand of mathematics, is often a discovery of a relation that is a true property of the one world.  

There is no necessity to limit the study of mathematical systems to those that elaborate these four core concepts.  But those that do display a vast richness of consequences and interconnections exactly because they are elaborations of core concepts that come from nature.  

One may then even define mathematics as the study of systems of evoked relationships inspired by observations of nature.  
Mathematics is then a system of objective facts, that is nonetheless timebound and open to unpredictable developments in the future

\section*{Conclusion}

If we give up the idea that there is a mathematical object existing in a timeless Platonic realm which is isomorphic to the history of the universe, we still have to explain why mathematics is so effective in physics.  It will be sufficient to point to an interpretation of the use of mathematics in physics that is consistent with the view of mathematics just presented.   Here is one such interpretation: {\it mathematics is useful as providing models that summarize the content of records of past observations.}  When we test a theory 
we make and record observations of motions, which consist of values of observables that we represent as the coordinates of the configuration space of a system.  These records are static, in that once taken they do not change in time.  Or, more precisely, they may change, by being degraded or erased, but once they do they cease to function as records of past experiments.   They can be compared to a trajectory in the configuration space which, being a mathematical object, is also static.

We can propose that the main effectiveness of mathematics in physics consists of these kinds of correspondences between records of past observations  or, more precisely, patterns inherent in such records, and properties of mathematical objects that are constructed as representations of models of the evolution of such systems.  This view does not require either the postulation that physical reality is timeless or that mathematical objects exist in a separate timeless realm.  It is sufficient that records of past observations are static and that the properties of a mathematical object are, once evoked into existence by their invention, static.  Both the records and the mathematical objects are human constructions which are brought into existence by exercises of human will, neither has any transcendental existence.  Both are static, not in the sense of existing outside of time, but in the weak sense that once they come to exist, they don't change.   



In closing, I would like to mention two properties enjoyed by the physical universe which are not isomorphic to any property of a mathematical object.

\begin{enumerate}

\item{} In the real universe it is always some present moment, which is one of a succession of moments.  Properties off mathematical objects, once evoked, are true independent of time.

\item{}The universe exists apart from being evoked by the human imagination, while mathematical objects do not exist before and apart from being evoked by human imagination.  

\end{enumerate}

{\bf Acknowledgements.}   I would like to thank Roberto Mangabeira Unger for collaboration on this project and Marina Cortes and Henrique Gomes for discussions on related work.  This research was supported in part by Perimeter Institute for Theoretical Physics. Research at Perimeter Institute is supported by the Government of Canada through Industry Canada and by the Province of Ontario through the Ministry of Research and Innovation. This research was also partly supported by grants from NSERC, FQXi and the John Templeton Foundation.



\begin{thebibliography}{99}


\bibitem{MUH}Tegmark, Max. Our Mathematical Universe: My Quest for the Ultimate Nature of Reality. Random House LLC, 2014; "The mathematical universe." Foundations of Physics 38.2 (2008): 101-150.


\bibitem{TN}Lee Smolin,
{\it Temporal naturalism }  arXiv:1310.8539. Invited contribution for a special Issue of Studies in History and Philosophy of Modern Physics, on Time and Cosmology, edited by Emily Grosholz.  

\bibitem{SURT}Roberto Mangabeira Unger and Lee  Smolin,  {\bf The Singular Universe and the Reality of Time}, (Cambridge University Press, 2015)


\bibitem{TR}Lee Smolin, {\it Time Reborn},  April 2013, Houghton Mifflin Harcourt, Penguin and Random House Canada.  

\bibitem{RMU-FQXi}Roberto Mangabeira Unger,  {\it  A Mystery Demystified: The Connection between Mathematics and Physics.}

\bibitem{me-brick}Lee Smolin, ÒThe Culture of Science Divided Against ItselfÓ, Brick Magazine, Issue 88, 2012.

\bibitem{Jaron-gadget}Jaron Lanier, You are not a gadget. Random House, Inc., 2010.


\bibitem{GS}Galen Strawson,  {\it Real naturalism}, London Review of Books,  Vol. 35 No. 18 á 26 September 2013
pages 28-30 


\bibitem{BMazur}Mazur, Barry. "Mathematical Platonism and its opposites." Newsletter of the European Mathematical Society (2008): 19-21.


\end{thebibliography}
\end{document}